\documentclass[prb,aps,twocolumn,amsmath,amssymb,floatfix,superscriptaddress]{revtex4-1}

\usepackage{physics}
\usepackage{amsmath}
\usepackage[dvips]{graphicx}% Include figure files
\usepackage[dvips]{graphics}
\usepackage{dcolumn}% Align table columns on decimal point
\usepackage{bm}% bold math
\usepackage{xcolor}\usepackage{appendix}
\begin{document}

\preprint{APS/123-QED}

\title{\textcolor{blue}{Engineering insulator-metal transition in a class of decorated aperiodic lattices: a quantum dynamical study}}% Force line breaks with \\

\author{Arkajyoti Maity}
 
 \email{psam2182@iacs.res.in,arkajyoti1212@gmail.com}
\affiliation{%
 School of Physical Sciences,\\Indian Association for the Cultivation of Science,\\
 2A \&\ 2B Raja S C Mullick Road, Kolkata 700032, India. 
}%

\author{Arunava Chakrabarti}
 \email{arunava.physics@presiuniv.ac.in}
\affiliation{
Department of Physics\\
Presidency University \\
86/1, College Street, Kolkata 700073, India% with \\
}%
\date{\today}% It is always \today, today,

\begin{abstract}
We investigate the quantum dynamics of wave packets in a class of decorated lattices, both quasiperiodic and random, where a nominal quasi-one dimensionality is introduced at local levels, bringing in a deterministic or even random variations in the distribution of the coordination number throughout the system. We show that certain correlations in the numerical parameters of the system Hamiltonian can cause a drastic change in the dynamical  evolution of the wave packet, revealing a complete delocalization, independent of the energy of the travelling particle, even in the absence of any translational invariance. We use an  exact decimation of a selected subset of the degrees of freedom, and an analysis of the commutation of the $2 \times 2$ transfer matrices on a renormalized version of the parent systems within a tight binding framework . An in-depth analysis of the mean square displacement, temporal autocorrelation function and the inverse participation ratio establishes the gross change in the behaviour of the wave packet dynamics. The consequence is the occurrence of a parameter-driven insulator-metal transition over the full (or a major) range of the energy spectrum in each case. In certain cases, inclusion of an external magnetic flux enables us to control the transition. The observation is general, and, to our mind, can inspire experiments involving photonics or matter wave localization.
\end{abstract}

\maketitle

%\tableofcontents

\section{\label{sec:level1}INTRODUCTION\protect  }
The effect of uncorrelated disorder in condensed matter systems is manifestly observed in the suppression of wave propagation of any kind. While the pathbreaking idea of localization of single particle states was put forward by Anderson~\cite{anderson} in 1958 for a three dimensional system modelled on a lattice, the subsequent analyses in two dimensions~\cite{abrahams}, and in one dimensional lattice models~\cite{borland} confirmed the quantum interference induced, disorder driven exponential localization of single particle states. The effect is strongest in 1D, where the states get localized irrespective of the strength of disorder. The phenomenon of localization is ubiquitous, and is alive even after more than sixty years since its inception. Recent experiments on the localization in disordered photonics~\cite{wiersma,cao,storzer,lahini}, or in the cases of matter waves~\cite{aspect,roati} have 
 rekindled the interest in this ever active field of research.

Over the last three decades, exciting variations on the classic examples of Anderson localization were brought to the notice of the condensed matter community, mainly through the appearance of delocalized, 
 extended (but non-Bloch) single particle states, that resulted mainly out of a selection of certain special values of the  energy of the propagating excitation. The initiation into this field was through the work on the so called {\it random dimer model} (RDM)~\cite{dunlap,phillips} where, a special positional correlation between the constituents of a binary, disordered alloy was shown to lead to {\it unscattered} eigenfunctions corresponding to a special energy eigenvalue. The results were experimentally verified by measuring the conductivity of positionally correlated multilayered structures~\cite{francisco}, or through a study of the anomalous wave packet evolution in an optical waveguide experiment~\cite{naether}, to name a few. The canvas of such studies became richer with the inclusion of quasiperiodic lattices in one dimension~\cite{macia, arunava1,arunava2}, or a class of deterministic fractals~\cite{arunava3}. While in the cases of quasiperiodic lattices in one dimension, the presence of local, positionally correlated clusters at all scales of length~\cite{arunava1,arunava2} was shown to generate a countable infinity of extended, totally transparent single particle states, for the deterministic fractals the geometry and self-similarilty  of the lattice as a whole were argued to be responsible for having densely populated extended eigenfunctions in an otherwise fragmented energy spectrum~\cite{arunava3}.

In this context, its worth mentioning that, even stronger variants of  delocalization of single particle states were brought to the notice of condensed matter community in recent times. Certain geometrically disordered lattices, described by a tight binding Hamiltonian with off-diagonal (hopping) disorder, and with minimal quasi-one dimensionality were shown to give rise to an {\it absolutely continuous} spectrum spanning the entire range of the allowed energy eigenvalues (or, most of it) when the hopping integrals connecting the cells in the lattice bore a certain ratio between them~\cite{biplab1,biplab2}. This finding is rather surprising and unusual, as it sets a condition on the choice of the {\it numerical values} of the system parameters, independent of the energy of the travelling electron (as in the RDM case) to realize a transition from a complete localization to a completely diffusive behaviour of the single particle excitations. In some cases, a properly tuned external magnetic flux trapped in the constituent `cells' of the lattice structures triggered such a complete (or, almost complete) delocalization of the single particle states. The basic scheme was extended to study the spin polarized transport and spin filtering effect in a quasiperiodic arrangement of the `building blocks' of such geometrically disordered lattices, where an external magnetic flux played a key-role~\cite{amrita}. However, a study of a possible engineering of such complete, or quasi-complete delocalization, through a close scrutiny of the quantum dynamics of the wave packet is still lacking, and this is our primary motivation for the present work.
A study of the quantum dynamics (of an electron, say) in any lattice begins with the time evolution of an initially localized wave packet~\cite{katsanos,arias}. For translationally ordered, perfectly periodic systems, the electronic transport is completely ballistic. For disordered lattices the transport is not very clearly defined, and the time evolution may vary from showing a localized character, in randomly disordered lattices for example, to a diffusive or even super diffusive behaviour as seen in certain quasiperiodic systems with weak disorder~\cite{zhong,thiem1,thiem2,thiem3}. However, as a rule of thumb we expect systems with absolutely continuous spectra to give rise to a  ballistic transport. Such transport classifications are derived from the scaling properties of dynamical probes, such as the mean square displacement and temporal auto-correlation function, to be explained in the next sections.

In this communication we focus on the quantum dynamics of spinless fermions and scrutinize the possibility of engineering a parameter-driven localization-delocalization transition in a couple of geometrically disordered lattices. In the geometries addressed here, a nominal quasi-one dimensionality is introduced through a cluster of atoms side coupled to a selected subset of the lattice points. Such a construction brings in a variation in the local coordination numbers in the lattice. The `disorder' is introduced through a distribution of the nearest neighbor hopping integrals in the the tight binding Hamiltonian. It is thus a problem of off-diagonally disordered systems, coupled with an `additional' disorder that comes from the distribution of the cells constituting the lattice structure. The structures presented here are only two members of a wide variety of lattices where such an {\it engineered  delocalization} is possible. Thus we have, in our hands, a group of  serious candidates where a clean violation of the canonical case of Anderson localization is observed~\cite{biplab1,biplab2}.

We make an in-depth analysis of the quantum dynamics of an electron released in such a system by examining  the time evolution of a wave packet, the mean square diplacement (MSD) and the autocorrelation function.  Such a dynamical study, to our mind can be implemented in a laboratory experiment either using photonic lattices developed using the ultrafast laser technology~\cite{sebabrata}, or using  tailor-made  ultracold atomic  `artificial' crystal structures~\cite{roati}. 
%%%%%%%%%%%%%%%%%%%%%%%%%%%%%%%%%%%%%%%%%%%%%
\begin{figure}[ht]
\includegraphics[width=\columnwidth]{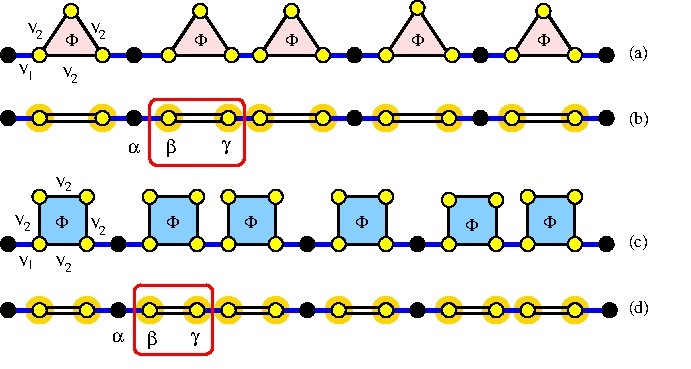}
\caption{(Color online.) The quasi-one dimensional triangle-dot (TD)  and the square-dot (SD) lattices ((a) and (c)). Both the geometries represent a quasiperiodic Fibonacci order. A renormalization decimation transformation maps both the lattices into effective one dimensional chains, as depicted in (b) and in (d). The trapped magnetic flux is written as $\Phi$.}
\label{lattice}
\end{figure}
%%%%%%%%%%%%%%%%%%%%%%%%%%%%%%%%%%%%%%%%%%%%%
In Section II we briefly describe the Hamiltonian and the quantitative probes adopted for investigating the quantum wave-packet dynamics,along with some known results. In Section III, we introduce the geometries considered for this work and give a brief discussion on the results obtained in the static case,for a better understanding of our results here.We then perform the actual dynamical study and present our results for the so called `triangle-dot' (TD) system, developed according to a quasiperiodic Fibonacci sequence. In the succeeding subsection, we work on  another geometry, viz, a square-dot (SD) array, but with a randomly disordered arrangement. We conclude by summarising our work in section IV, and discuss the prospects of any future work.

\section{\label{sec:level1}Dynamics of single particle systems: Measures of investigation and some results\protect  }
\subsection{\label{sec:level1}The  Hamiltonian\protect}
The dynamics of a particle in a lattice is governed by the tight binding  Hamiltonian:

\begin{equation}
H = \sum_{n}\epsilon _n\mid n \rangle\langle n\mid + \sum_{\left \langle n,m\right \rangle}V_{nm}\mid n   \rangle \langle m\mid     
\end{equation}
where, the $\ket{n}$ is the Wannier orbital at the $n$-th site, $\epsilon_n$ is the 
corresponding on-site potential, and $V_{nm}$ is the nearest neighbor hopping integral. The amplitude of the wave function of an electron at the $n$-th site 
satisfies the time independent Schr\"{o}dinger equation, written equivalently in terms of a set of difference equations as, 
\begin{equation}
 V_{n,n+1}\psi_{n+1} +V_{n-1,n}\psi_{n-1}+(\epsilon_{n}-E)\psi_{n} = 0
 \label{diff}
 \end{equation}
where E is the energy eigenvalue. 

A wave packet
initially released at any site at time $t=0$, evolves in time following the expression~\cite{katsanos}, 
\begin{eqnarray}
    \psi(t) & = & \exp{-iHt/\hbar}\psi(0) \nonumber \\
   &  = & \sum_{k}\exp{-iE_{k}t/\hbar}c_{k}(0)\psi_{k}(0)
\label{wavepacket}
\end{eqnarray} 
where $c_k(0)$ are the initial coeffecients of the wavefunction.  Eq.~\eqref{wavepacket} is going to be studied in depth in order to figure out the quantum dynamics of an electron in a class of decorated, aperiodically ordered cells.

\subsection{\label{sec:level1}Quantifying the dynamics \protect}
The single particle dynamics is conveniently described in terms of the mean square displacement (MSD). If a wave-packet, initially localized at a particular site $n_{0}$ is allowed to evolve in time, we can estimate it's spread by evaluating the MSD, which is the variance measure, viz, 
\begin{equation}
    \sigma^{2}(t)=\sum_{n}(n-n_{0})^{2} (\psi_{n})^{2}
\end{equation}
where $n$ is refers to the lattice site-index.
In a perfectly periodic system with an absolutely continuous spectrum, $\sigma^{2}(t)$ has an asymptotic dependence on time~\cite{katsanos}, in the form of a power law, viz,  $\sigma^{2}(t) \sim  t^{2}$. 
This represents the ballistic motion. In general, we expect the $\sigma^{2}(t)$ to go as a power law with time: 
~$t^{\mu}$. \textcolor{blue}{Localization corresponds to $\mu < 1$, while $\mu = 1$, $\mu > 1$ and $\mu =2$ correspond to an ordinary diffusion, superdiffusion and ballistic motion respectively}. 

Another well implemented measure is the temporal autocorrelation function (TAF), which is a time averaged representation the return probability of the wave function back to it's initial site of release.
The TAF  $C(t)$ is defined as, 
\begin{equation}
 C(t)=\frac{1}{t}\int_{0}^{t}\left | \psi_{n_{0}} \right |^{2}dt   
\end{equation}
where $n_{0}$ is the initial site of release.
The decay of $C(t)$ for large times is expected
to obey the power-law:
$C(t)\sim t^{- \delta}$.
The exponent $\delta$ approaches unity for long times in completely periodic systems, while it remains static in extreme disordered cases.
A higher fall rate is attributed to a metal-like behaviour.

The final measure which we implement is the inverse participation ratio (IPR). It is defined as the integral (or a discrete sum for finite system size) over the square of the density in some space, e.g., real space, momentum space, or even phase space, where the densities would be the square of the real space wave function $\left | \psi(x) \right |^{2}$, the square of the momentum space wave function $\left | \tilde{\psi}(k) \right |^{2}$ or some phase space density like the Husimi distribution.
For discrete systems the IPR is given by:
\begin{equation}
  \mathcal{I} =  \sum_{n}\left | \psi_{n} \right |^{4}
\end{equation}
For a system of $N$ sites, the maximum ($= 1$) of the IPR corresponds to a fully localized state whereas, the minimum corresponds to delocalization for which $\psi(x) = \frac{1}{\sqrt{N}}$ and the ratio comes out as $\frac{1}{N}$.

In what follows, along with  the MSD or TAF, we also carefully investigate the change in the IPR, in particular how its value decreases for our specified class of decorated lattice models where a numerical correlation between the values of the hopping integrals has been known to trigger a complete change of the character of the single particle states\cite{biplab1,biplab2}, going from an Anderson localized shape to a perfectly extended, fully transmitting one. 
%%%%%%%%%%%%%%%%%%%%%%%%%%%%%%%%%%%%%%%%%%%%%%%%%%%%%%%%
\begin{figure}[ht]
\centering
\includegraphics[width=0.9\columnwidth]{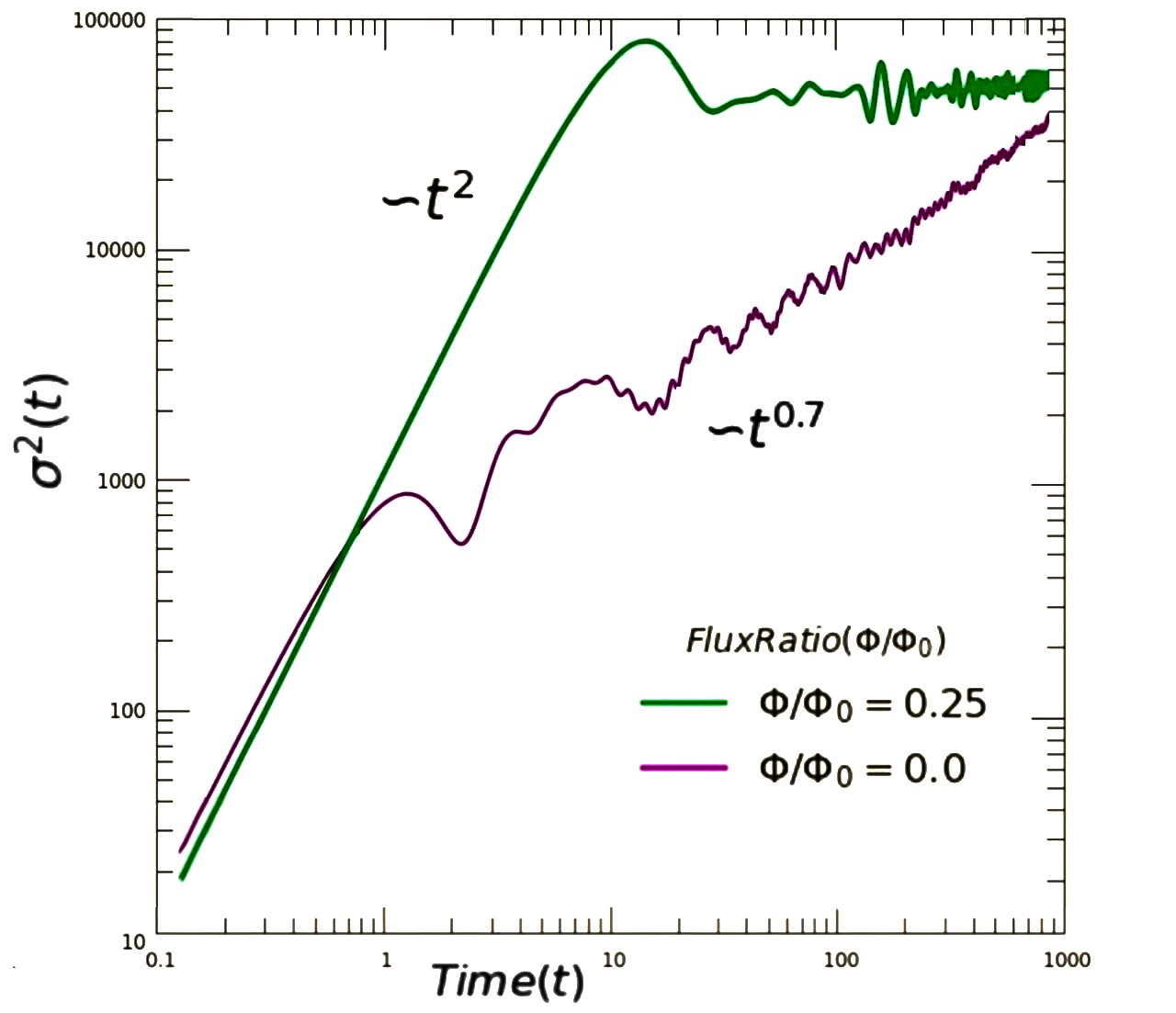}
 \caption{Mean square displacement for a 843-site triangle-dot system where $V_1/V_2 = \sqrt{2}$, and $\Phi=0$ (violet) and $\Phi=\Phi_0/4$ (green) respectively. \textcolor{blue}{The sites are numbered from the left on the original (undecimated) lattice.}}
\label{msd}
\end{figure}
%%%%%%%%%%%%%%%%%%%%%%%%%%%%%%%%%%%%%%%%%%%%%%%%%%%%%%%%

\section{\label{sec:level 2}Dynamics of some aperiodic systems and breakdown of Anderson localization \protect  }

\subsection{\label{sec:level1}Triangle-Dot system\protect}
The first system chosen, is an array of triangles and dots as shown in Fig.~\ref{lattice}(a). The vertices of a triangle are the `atomic' sites that are connected to their nearest neighbours along the arms of the triangle by a hopping integral $V_{nm} \equiv V_2$. All other hopping integrals are assigned a value $V_1$. The triangular plaquettes, and the isolated dots (solid black circles in Fig.~\ref{lattice}(a)) are chosen to be distributed following a quasiperiodic Fibonacci sequence, though for the discussion that follows, a random arrangement of such building blocks would make no difference in the conclusion. A constant magnetic flux $\Phi$ is trapped within the triangular cavities. This results in a Peierls'  phase factor tagged with the hopping amplitude along the arms of each triangle, breaking the time reversal symmetry locally, along each arm. We choose a gauge so that, along each arm of a triangle the hopping is, $V_2 \exp (\pm i 2\pi \Phi/3\Phi_0)$. Here $\Phi_0=hc/e$ is the fundamental flux quantum.

\subsubsection{A parameter-tuned delocalization of eigenstates}

Let us look at Fig.~\ref{lattice}(a). Using Eq.~\eqref{diff} it is easy to decimate out the top vertices of every triangle, leading to a linear chain of atomic sites, comprising the original, black dots (Fig.~\ref{lattice}(b)), now named $\alpha$, and pairs of sites, named $\beta$ and $\gamma$, lying at the base of any triangle. The self energies of the $\beta$ and $\gamma$ sites are renormalized, along with the hopping integrals connecting them~\cite{biplab1}. The black sites $\alpha$ retain their original on-site potential $\epsilon$, while the renormalized values of the $\beta$ and the $\gamma$ sites are given by,
\begin{equation}
    \epsilon_\beta = \epsilon_\gamma = \epsilon + \frac{V_2^2}{E-\epsilon}
\label{onsiterg}
\end{equation}
The {\it effective} hopping integral connecting the $\beta\gamma$ pair of sites now reads, $V_{\beta\gamma}=V_{\gamma\beta}^\ast = \tau \exp (i\xi)$, where,
\\
\begin{equation}
\begin{aligned}
\tau & = \sqrt{V_2^2 + \frac{V_2^4}{(E-\epsilon)^2} + \frac{V_2^3}{(E-\epsilon)} \cos 3\theta \nonumber} \\
\tan \xi & =  \frac{(E-\epsilon) V_2 \sin \theta - V_2^2 \sin 2\theta}
{(E-\epsilon) V_2 \cos \theta + V_2^2 \cos 2\theta}
\end{aligned}
\end{equation}
where, $\theta=2 \pi \Phi/(3 \Phi_0)$. 
Now, for the sake of completeness, and an appreciation of the results of the quantum dynamical study, we summarise the essential results, already discussed elsewhere~\cite{biplab1} in details, below.

The amplitude of the wavefunction, and the energy spectrum of a sequence (Fibonacci or, any other arbitrary arrangement) of $\alpha$ and $\beta\gamma$ can be conveniently obtained through
a product of $2 \times 2$ {\it transfer matrices}, viz, $\mathcal{M_\alpha}$, and $\mathcal{M_{\gamma\beta}} \equiv \mathcal{M_\gamma} \mathcal{ M_\beta}$~\cite{kohmoto}, which in our case, are given by,
\newline
\newline
\begin{math}
\mathcal{M_\alpha}=  \left [{\begin{array}{cc}
    (E-\epsilon)/V_1 & -1 \\
    1 & 0 \\
    \end{array}} \right ] , \quad\hfill 
    \mathcal{M_{\gamma\beta}}=  \left [{\begin{array}{cc}
    m_{11} & m_{12}\\
    m_{21} & m_{22} \\
    \end{array}} \right ]
%    \end{align} 
\end{math}

The elements of the matrix $M_{\gamma\beta}$ are given by, 
\begin{eqnarray}
m_{11} & = & e^{-i\xi} \left[ \frac{(E-\epsilon_\gamma) (E-\epsilon_\beta)}{V_1\tau}-\frac{\tau}{V_1} \right ] \nonumber \\
m_{12} & = & -\frac{(E-\epsilon_\gamma)}{\tau} e^{-i\xi} \nonumber \\
m_{21} & = & -\frac{(E-\epsilon_\beta)}{\tau} e^{-i\xi} \nonumber \\
m_{22} & = & -\frac{V_1}{\tau} e^{-i\xi}
\end{eqnarray}
It can now be easily verified that~\cite{biplab1}, the commutator $[\mathcal{M_\alpha}, \mathcal{M_{\gamma\beta}}] = 0$ if we set $V_2=V_1/\sqrt{2}$, and $\Phi=\Phi_0/4$. This commutator vanishes {\it independent of the energy} $E$ of the propagating particle, and lies at the heart of a completely non trivial variation of the Anderson localization, going beyond the RDM's where, an extended state came up only at certain special values of the energy. 

The `energy independent' commutation of $M_\alpha$ and $M_{\gamma\beta}$ implies that the order of the arrangement of the clusters $\beta\gamma$ and $\alpha$ is {\it immaterial} now, for the entire spectral range, and therefore any disordered, infinitely long array of the triangles and dots thus becomes indistinguishable from a perfectly periodic array of the same, as far as the nature of the single particle states are concerned (in fact, the energy bands also merge, as has been shown before~\cite{biplab1}). Absolutely continuous energy bands (at least subbands, spanning most of the spectrum), and extended Bloch-like eigenstates characterize the system as soon as the {\it resonance} condition is satisfied. Of particular interest is the observation that, if we set $V_2=V_1/\sqrt{2}$ at the very outset, that is, we `fix' our lattice at the beginning, then we can trigger this transition from a localized (for an aperiodic array) to an extended (for a periodic array) nature of the eigenstates, that is, an insulating to metallic phase by tuning an {\it external parameter}, which is the magnetic field in the present case.

\subsubsection{Quantum dynamics: The Mean Square Displacement}
Here we try to see how such `resonance' conditions influence the electron dynamics. We use the full Hamiltonian now, without `decimating' any sites. In the presence of a magnetic field piercing the triangular cells, the hopping along the arms of the triangle are given by $V_2~\exp[\pm i 2\pi\Phi/3\Phi_0]$, while the hopping along the backbone remains $V_1$. In this particular subsection, we choose a Fibonacci array of triangles and dots, without any loss of generality.

Denoting the Wannier orbital at the $i$th site as $\ket{i}$, we find the amplitudes $\psi_{i}(0) \equiv \ket{i}_{n_{0}}$ of each such orbital at time $t=0$. The wavepacket is released at a site (chosen arbitrarily) marked $n_0$. It is thus a delta function concentrated at the site $n_0$, at time $t=0$. 
At any later time $t$, the coefficients evolve through the Hamiltonian and as per equation (6)  we superpose the solutions to form the complete wavefunction. 
The mean square displacement (MSD), as already detailed out in Section II, has been examined for the triangle-dot system, and the results are shown in Fig.~\ref{msd}. 
%%%%%%%%%%%%%%%%%%%%%%%%%%%%%%%%%%%%%%%%%%%%%%%%%%%%%%%%%%
\begin{figure}
\includegraphics[width=0.9\columnwidth]{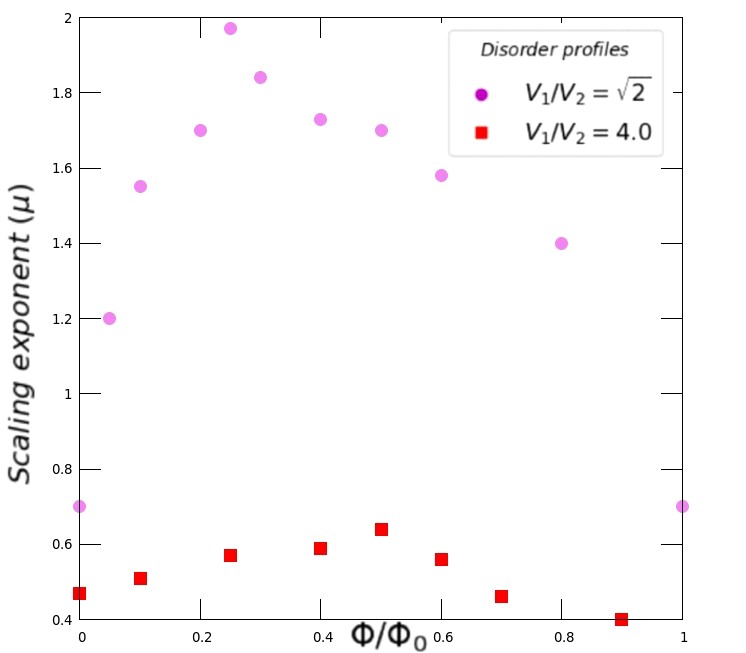}
 \caption{Scaling exponent ($\mu$) plotted against the magnetic flux ratio. The red data points are for the $V_1/V_2 = 4$, and the violet dots represent the $V_1/V_2=\sqrt{2}$ condition. We see that a magnetic field actually changes the scaling exponent considerably in the violet data set and indeed we get ballistic transport in the resonance condition. }
\label{scaling}
\end{figure}
%%%%%%%%%%%%%%%%%%%%%%%%%%%%%%%%%%%%%%%%%%%%%%%%%%%%%%%%%%
To make our MSD arguments robust, all plots shown are averaged over $20$ simulations where the wavefunction begins to time-evolve from a chosen (arbitrary) site in a finite Fibonacci segment having a total of $843$ sites, including the dots and the triangles.

As is evident in Fig.~\ref{msd}, both for $\Phi/\Phi_0=0$ and $\Phi/\Phi_0 \ne 0$ the  wavepacket begins to expand, scaling similarly. However, we only care about it's nature at exponentially large time frames. In Fig.~\ref{msd}, with resonance condition incorporated, the ballistic nature of the motion of the wavepacket is apparent at large times. In the off-resonance case with $\Phi=0$ (the deep violet curve), the wavepacket needs to `spend' some time to smell the disorder (or so to say the lack of any order equivalent to that in a periodic chain) and thus follows the green curve upto a certain finite time scale, after which the disordered environment takes over and the MSD exhibits the typical log periodic oscillations before converging to certain value. The convergence is essentially a manifestation of dealing with a finite system. For a finite sized lattice, there will be an upperbound to the MSD, as suggested by the Cauchy-Schwartz inequality, viz, 
\begin{equation}
     \sum_{n}\psi_{n}^2(n-n_{0})^2\leqslant \left (\sum_{n}\psi_{n}^2 \right )\left ( \sum _{n} (n-n_{0})^2\right )
     \end{equation}
That is, equivalently one can write, 
 \begin{equation}
 \sum_{n}\psi_{n}^2(n-n_{0})^2\leqslant\left ( \sum _{n} (n-n_{0})^2\right )
\end{equation}
The upperbound of the mean square displacement, as obtained above,  allows us to set up an arguably more accurate measure of `large time' by considering only the time during which the MSD grows from a pre-chosen value to close to the upperbound.

In Fig.~\ref{scaling} the MSD scaling is shown for two cases. In this figure, the influence of the external magnetic field, or rather the impact of the special flux ratio of $\Phi/\Phi_0=1/4$ is distinctly seen. With the on-site potential chosen to be zero always, we study the variation of the scaling exponent $\mu$ as the flux trapped in the triangular plaquettes is made to vary between zero and one flux quantum. The ratio of the hopping amplitudes are chosen to be  $V_1/V_2=4$ (red dots) and $V_1/V_2= \sqrt{2}$ (violet dots) respectively. In this scaling analysis, 
we choose to investigate the time during which MSD increases from $100$ to near about $10^5$ - a jump of three orders of magnitude.

When $\Phi=0$, we find that at 'large' times, the MSD ($\sigma^2(t))$ scales in time with an exponent $\mu$ is equal to $0.7$, while it grows to $1.92$ as we tune the flux to the value $\Phi=\Phi_{0}/4$. This is indicative of a sharp transition where the triangle-dot system switches over from a localized, insulating character to one showing ballistic transport, and hence a metallic behaviour. 

Interestingly, for a choice of a large $V_1:V_2$ ratio (the red dots) the magnetic field has little or practically no effect on the transport. This is clearly seen in Fig.~\ref{scaling} , in the distribution of the red dots. The quantum states remain localized ($0.46<\mu<0.6$) as we vary the flux ratio from zero to unity. With $V_1/V_2=\sqrt{2}$, repeated and extensive simulations reveal that, as the flux ratio is increased gradually, the return probability (and hence, the transport) quickly becomes diffusive, superdiffusive and approaches ballistic nature as $\Phi/\Phi_0 \rightarrow 1/4$. Interestingly, the system scales down to the localized character again beyond the special value of $\Phi=\Phi_0/4$ at a much slower rate. This implies that, even in the `off resonance' condition, one can still observe macroscopically large patches of {\it extended} eigenstates forming mini-bands, and yielding very good transport. This dynamical result thus corroborates the findings in the static cases~\cite{biplab1}.

\subsubsection{Temporal autocorrelation function:}
In this part we study the temporal autocorrelation function which has been duly defined in Section II. This measure is local in the sense that, it depends only on the return probability of the wavefunction and it's change with time.
%%%%%%%%%%%%%%%%%%%%%%%%%%%%%%%%%%%%%%%%%%%%%%%%%%%%%%%%%%%%%
\begin{figure}[ht]
\centering
\includegraphics[width=0.9\columnwidth]{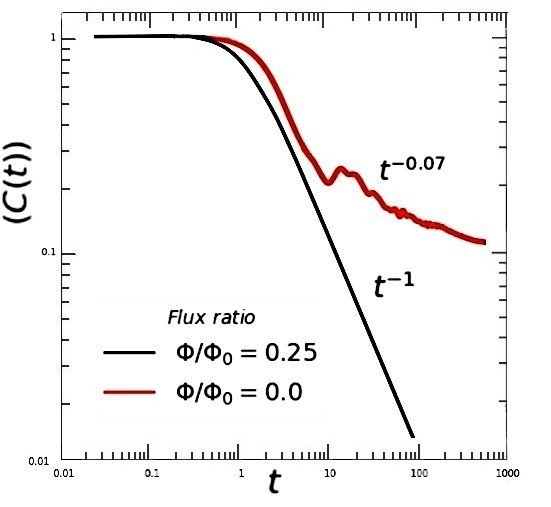}
 \caption{Temporal autocorrelation function for a 843-site long Fibonacci TD lattice where $V_1/V_2=\sqrt{2}$ in magnetic field free (red) case and in the resonance condition (black). The scaling exponents at large times are shown above the curves as the powers of $t$ (time).}
\label{auto}
\end{figure}
%%%%%%%%%%%%%%%%%%%%%%%%%%%%%%%%%%%%%%%%%%%%%%%%%%%%%%%%%%%%%

We work with the previously described Fibonacci sequence of triangles and dots, the total number of sites being 843, and we use a hard wall boundary condition. The on-site potential and the hopping integrals are described in the same way as before. We keep $V_1/V_2=\sqrt{2}$, and study the scaling of the auto correlation function at large times. In the flux free case, we are `off resonance' and the quasiperiodic arrangement of the building blocks keeps the eigenfunctions non-extended. This is manifestly observed in Fig.~\ref{auto}. The autocorrelation function seems to be pinned at $C(t)=1$ for some time and then starts dropping off. However the drop is not sharp at all, and as time passes it slows down even more, leading to a power-law decay at larger times. The `fit' to this drop at large time scales has been found to be $C(t) \sim t^{-0.07}$. 

On the other hand, with a magnetic flux $\Phi=\Phi_0/4$ trapped in every triangle, we observe a prominent sharp fall in the auto-correlation. This is depicted by the black line in Fig.~\ref{auto}. The large time behaviour of the autocorrelation function now reveals a scaling law $C(t) \sim t^{-1}$, and  implies a ballistic transport - a distinctive feature of perfectly periodic systems.

\begin{figure}[ht]
\centering
\includegraphics[width=0.9\columnwidth] {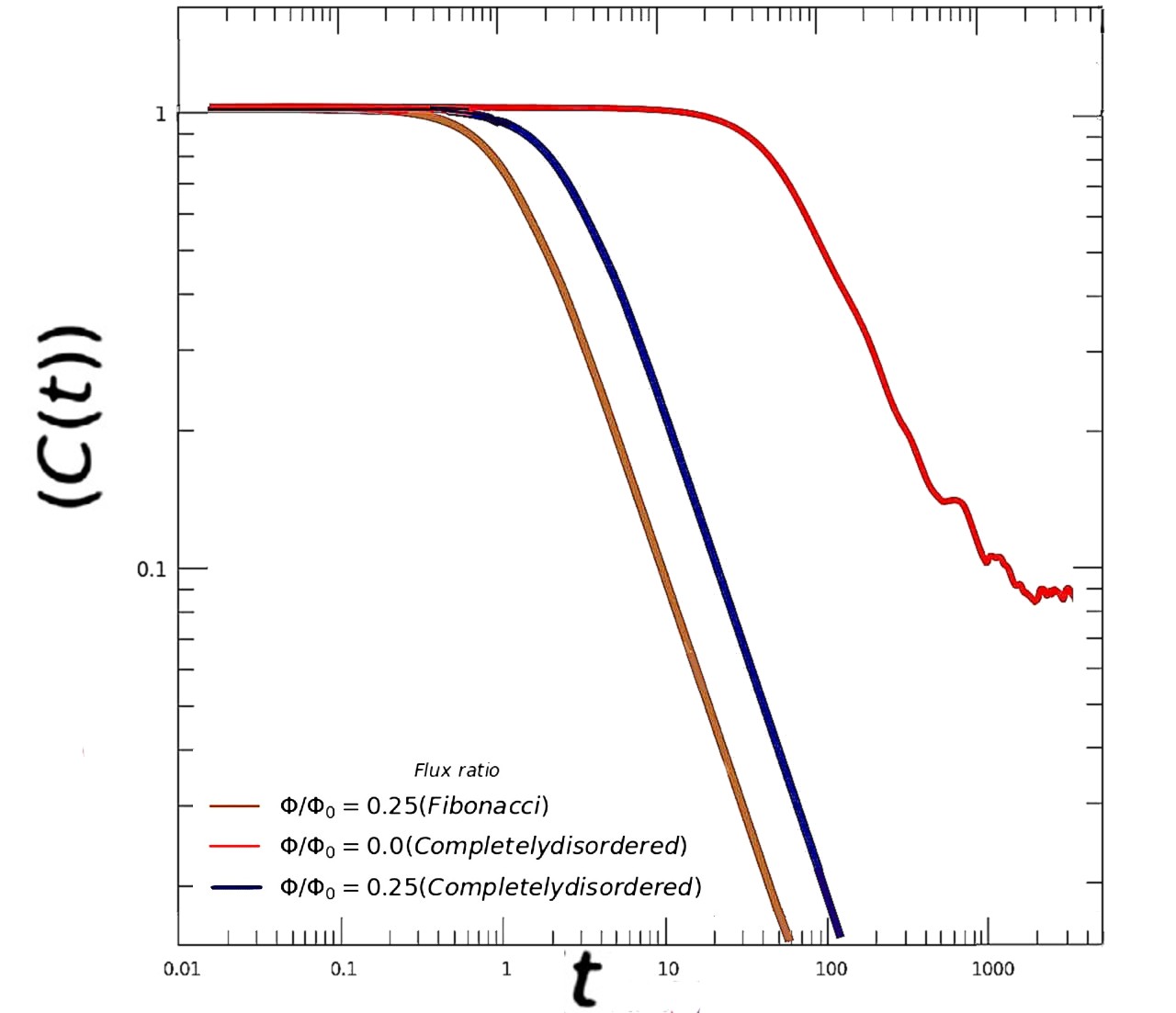}
\caption{Temporal autocorrelation function for the different configurations of the 843-site long TD geometry in flux free and resonance flux conditions. For every graph we have set $V_1/V_2=\sqrt{2}$.}
\label{disorder}
\end{figure}

We have repeated the numerical experiment with  completely disordered arrangements of the triangles and the dots (twenty different frozen-in disordered arrangements). The wave packet is released from different sites, and the autocorrelation function is averaged over all such configurations. In Fig.~\ref{disorder} we present this case. Interestingly, for the completely disordered system, in absence of flux, the autocorrelation function remains pinned at unity for a much longer time, and eventually starts decreasing, following a power law $C(t) \sim t^{-0.1}$, as depicted by the red line in Fig.~\ref{disorder}. A non-zero flux, particularly, $\Phi=\Phi_0/4$ makes $C(t)$ for both the Fibonacci and the randomly disordered array of triangles and dots drop off by two orders of magnitude. In Fig.~\ref{disorder} we show both these results by the brown (Fibonacci case) and blue (disordered case) respectively. The pinning of $C(t)$ at a value equal to unity implies a trend of retaining strong concentration of probability at the site of release. This reflects the expected localization. The long time decline in $C(t)$, seen even in the off-resonance cases (the red line in Fig.~\ref{disorder}) is of course, the finite size effect. 
The long time behaviour of $C(t)$ remains the same as soon as the flux through each triangle is set as $\Phi_0/4$. The geometry-independence of the result is in conformity with the energy-independent commutation of the transfer matrices, as discussed before.

Hence, we claim with confidence that the introduction of our chosen value of magnetic flux inside the triangular plaquettes has indeed transformed the physics of the system from that of a quasiperiodic (or, disordered) one to that of an {\it effective} periodic array of the triangles and dots, despite the spatial inhomogeneity in their distribution, and that, in terms of the long time behaviour of $C(t)$, a disordered array of triangles and dots becomes indistinguishable from a quasiperiodic Fibonacci (or, for that matter, any kind of aperiodic arrangement) array of the same. The interesting aspect in this study is that, with a given choice of $V_1/V_2=\sqrt{2}$ (which actually defines the system once and for all), one can trigger a metallic behaviour in an otherwise insulating system, by tuning an external agent - the magnetic field in this case.
%%%%%%%%%%%%%%%%%%%%%%%%%%%%%%%%%%%%%%%%%%%%%%%%%%%%%%%%%%%%%
%%%%%%%%%%%%%%%%%%%%%%%%%%%%%%%%%%%%%%%%%%%%%%%%%%%%%%%%%%%%%

\subsubsection{Inverse Participation Ratio: The random disorder case}
The inverse participation ratio (IPR) has already been described in detail in Section II.
Here we study how the IPR for a randomly disordered array of triangles and dots changes with time. Although there is no steadfast scaling relation of the IPR defined to demarcate the nature of electron transport, we qualitatively understand that an IPR having a value close to unity for long time scale will imply a localized particle. On the other hand,  if it drops sharply down to it's minimum of $1/N$, $N$ being the number of sites, an inference of delocalization can be drawn.

For the IPR calculations, we keep $V_1/V_2$=$\sqrt{2}$ and then average over $20$ disorder configurations. We then study two cases, viz, when $\Phi=0$ and when $\Phi=\Phi_{0}/4$. The results are shown in Fig.6

\begin{figure}[ht]
\centering
\includegraphics[width=0.9\columnwidth]{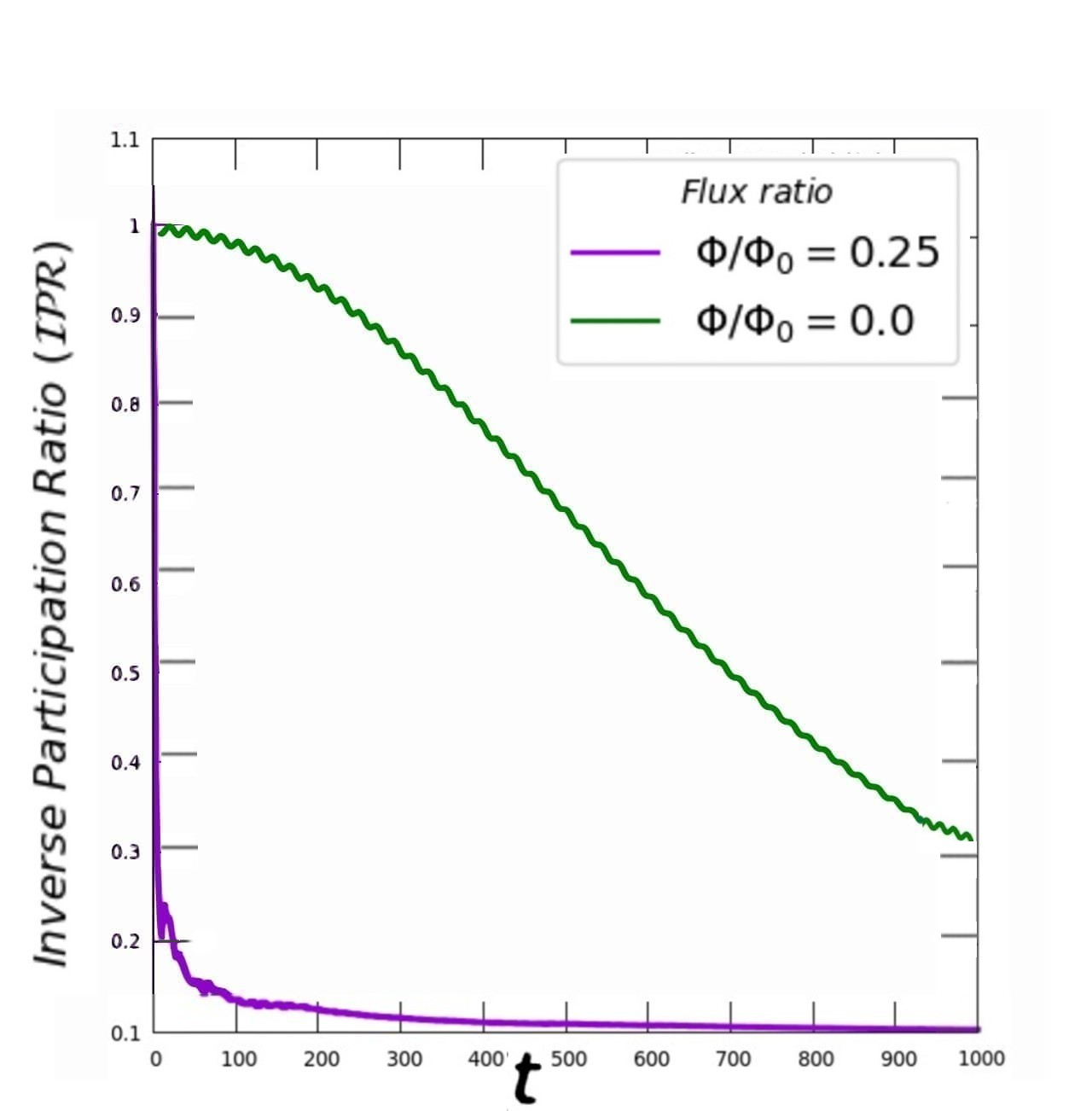}
 \caption{The IPR, averaged over $20$ disordered arrangements of triangles and dots. The total number of sites is now $850$, including dots and the triangle- vertices. $V_1/V_2=\sqrt{2}$.}
\label{ipr}
\end{figure}
We observe that in the zero field condition, the wavepacket delocalizes at a much slower rate, in comparison to the resonant case. The IPR with the flux $\Phi$ set at $\Phi_0/4$ settles at an order of magnitude lower value, and saturates to the lower bound at a time when the $\Phi=0$  case still clings to a value very close to unity. Even at $t=1000$, the IPR for the flux free case is an order of magnitude higher. Its subsequent drop is a result of the finite size of the system. This result corroborates our observations in the MSD and the temporal auto- correlation function studies.
\subsection{The Square-Dot system}

Figs.~\ref{lattice} (c) and (d) show our second geometry of consideration, and its renormalization to an effective one dimensional chain respectively. We have shown a Fibonacci pattern in the arrangement of the squares and the dots in Fig.~\ref{lattice}(c) just to illustrate the construction, but in this discussion we go for a completely random arrangement of squares and dots (SD). In any case, the portion of the lattice depicted in Fig.~\ref{lattice}(c) can as well form a part of a disordered arrangement. 

Just like in the previously discussed TD  system, here too the onsite potentials are chosen to be zero and the hopping integral is $V_2$ along the arms of a square, and $V_1$ everywhere else along the backbone. A flux, as before is made to get trapped inside a square, and our choice of a suitable gauge tags every hopping $V_2$ with a Peierls' phase factor, 
\begin{displaymath}
   V_2 \rightarrow V_2 \exp{\pm{i\frac{2\pi}{4}\frac{\Phi}{\ \Phi_{0}}}}  
\end{displaymath}

We can easily decimate out the top vertices of every square plaquette to map the system to an effective linear chain with modified onsite potentials and hoppings. Using the same nomenclature as described in the TD system, we have,
%%%%%%%%%%%%%%%%%%%%%%%%%%%%%
\begin{eqnarray}
\epsilon_\beta &  = & \epsilon+\frac{{V_2}^2}{\Delta} \nonumber \\
V_{\beta\gamma} & = & \tau e^{i\xi}
\label{rgsquare}
\end{eqnarray}
%%%%%%%%%%%%%%%%%%%%%%%%%%%%%
where, 
%%%%%%%%%%%%%%%%%%%%%%%%%%%%%
\begin{eqnarray}
\Delta & = & (E-\epsilon)^2 - V_2^2 \nonumber \\
\tau & = & V_2 \sqrt{\frac{V_2^2}{\Delta^2} + 2 \frac{V_2}{\Delta} \cos~4\theta + 1} 
\nonumber \\ 
\xi & = & \tan^{-1} \frac{\Delta \sin~\theta - V_2 \sin~3\theta}{
\Delta \cos~\theta + V_2\cos~3\theta}
\end{eqnarray} 
%%%%%%%%%%%%%%%%%%%%%%%%%%%%%
Here, $\theta=2\pi\Phi/4\Phi_0$. Obviously, $\epsilon_\gamma=\epsilon_\beta$, and $V_{\gamma\beta}=V_{\beta\gamma}^{\ast}$. 
This allows us to use the same transfer matrix formalism with the same shapes of the transfer matrices, and the conditions for a completely {\it energy-independent} commutation of all  the transfer matrices turn out to be the same, viz, $V_1/V_2 = \sqrt{2}$, and $\Phi=\Phi_0/4$. 
These conditions have been tested to be true for a much wider variety of geometrical shapes and arrangements, such as an array of an $n$-sided polygon and the dots sitting on a line, and a flux threading the polygons. In each case the `resonance condition' remains the same. This implies a subtle {\it universality} hidden in such decorated lattices. 

We now proceed to show   whether a single particle transport in such a spatially disordered system can be made to mimic that of a periodic system by hitting such `universal' resonance conditions. The IPR and temporal autocorrelation functions are once again employed as our tools to study the transport. The algorithm for evolution of the wavefunction, initially localized at any site, is done in the same line, as described in Section II.
%%%%%%%%%%%%%%%%%%%%%%%%%%%%%%%%%%%%%%%%%%%%%%%%
 \begin{figure}[ht]
\includegraphics[width=0.9\columnwidth]{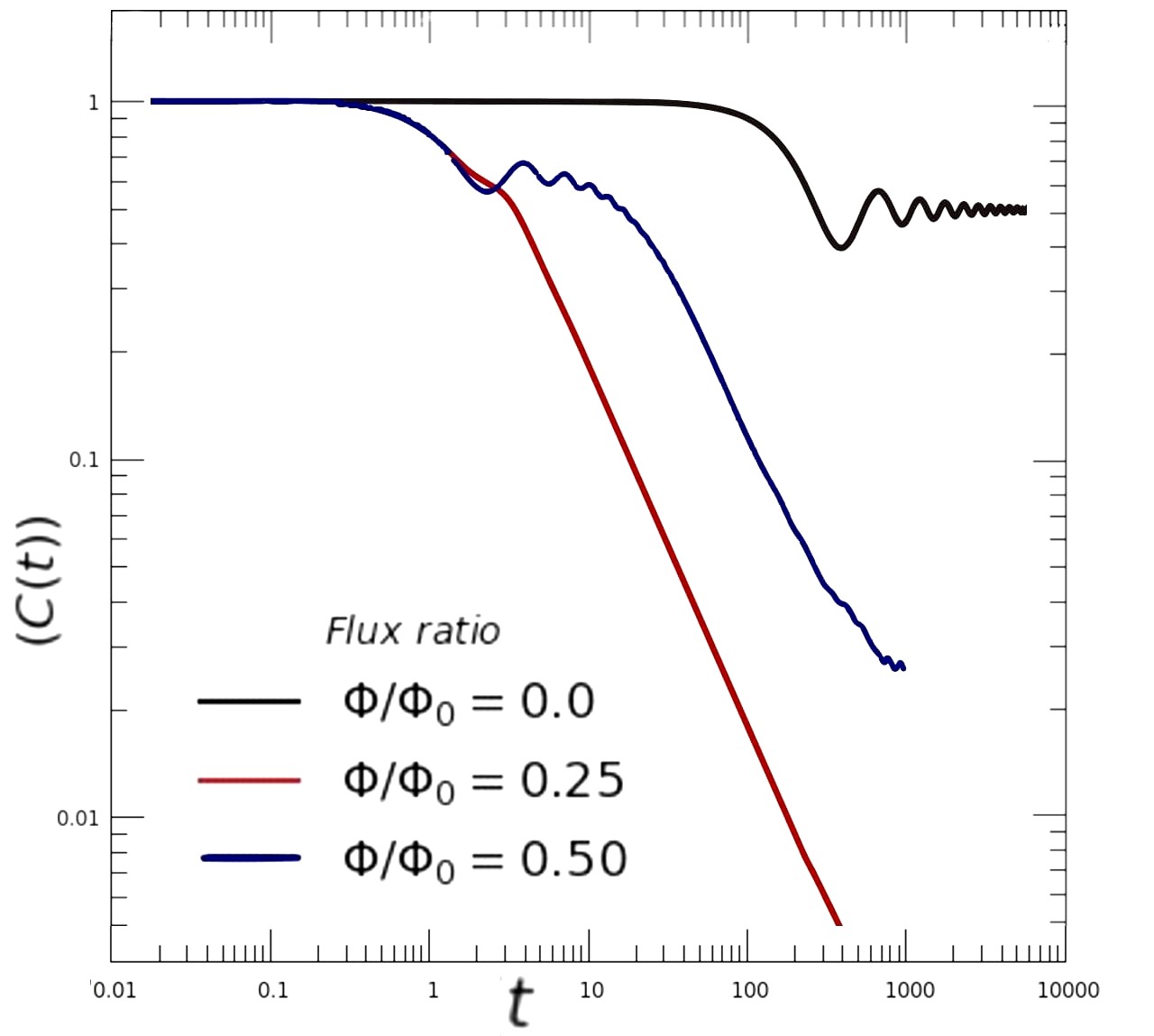}
\caption{Temporal autocorrelation profile averaged over computer generated maximally disordered arrangements of 1066 sites in SD configuration, for different flux values. The ratio  $V_1/V_2=\sqrt{2}$ here.}
\label{sdcorrelation}
\end{figure}
%%%%%%%%%%%%%%%%%%%%%%%%%%%%%%%%%%%%%%%%%%%%%%%%
We study the temporal autocorrelation function for the condition $V_1/V_2 = \sqrt{2}$, and use the flux ratio as our control parameter. As shown in Fig.~\ref{sdcorrelation}, for the zero field case, the autocorrelation is almost constant with time even for very large times (black line), only shows some ripples and a drop (though not to an appreciably low value), in the plateau for $t \ge 1000$, which we attribute to the finite size of the system again.  The signature is that of a strong localization of the eigenstates. As we had hoped, for the resonance condition, when the flux $\Phi=\Phi_0/4$, the correlation function drops rapidly towards zero (maroon line). The drop shows a scaling with scaling $C(t) \sim t^{-\delta}$, with $\delta=1$, which is the clear signature of a  ballistic transport. 

Interestingly, even with a $100 \%$ deviation from the ideal resonance flux value of $\Phi_0/4$, the autocorrelation tends to decrease pretty quickly as time passes on (navy blue line), revealing that even when the transfer matrices do not {\it exactly commute} for any energy, the strictly localized character of the eigenstates over the entire range of the spectrum gets disturbed, and that the spectrum of the any geometrically disordered SD system can still offer minibands of extended states, with the transport being reasonably good over these energy subbands.
%%%%%%%%%%%%%%%%%%%%%%%%%%%%%%%%%%%%%%%%%%%%%%%
 \begin{figure}[ht]
 \centering
\includegraphics[width=0.9\columnwidth]{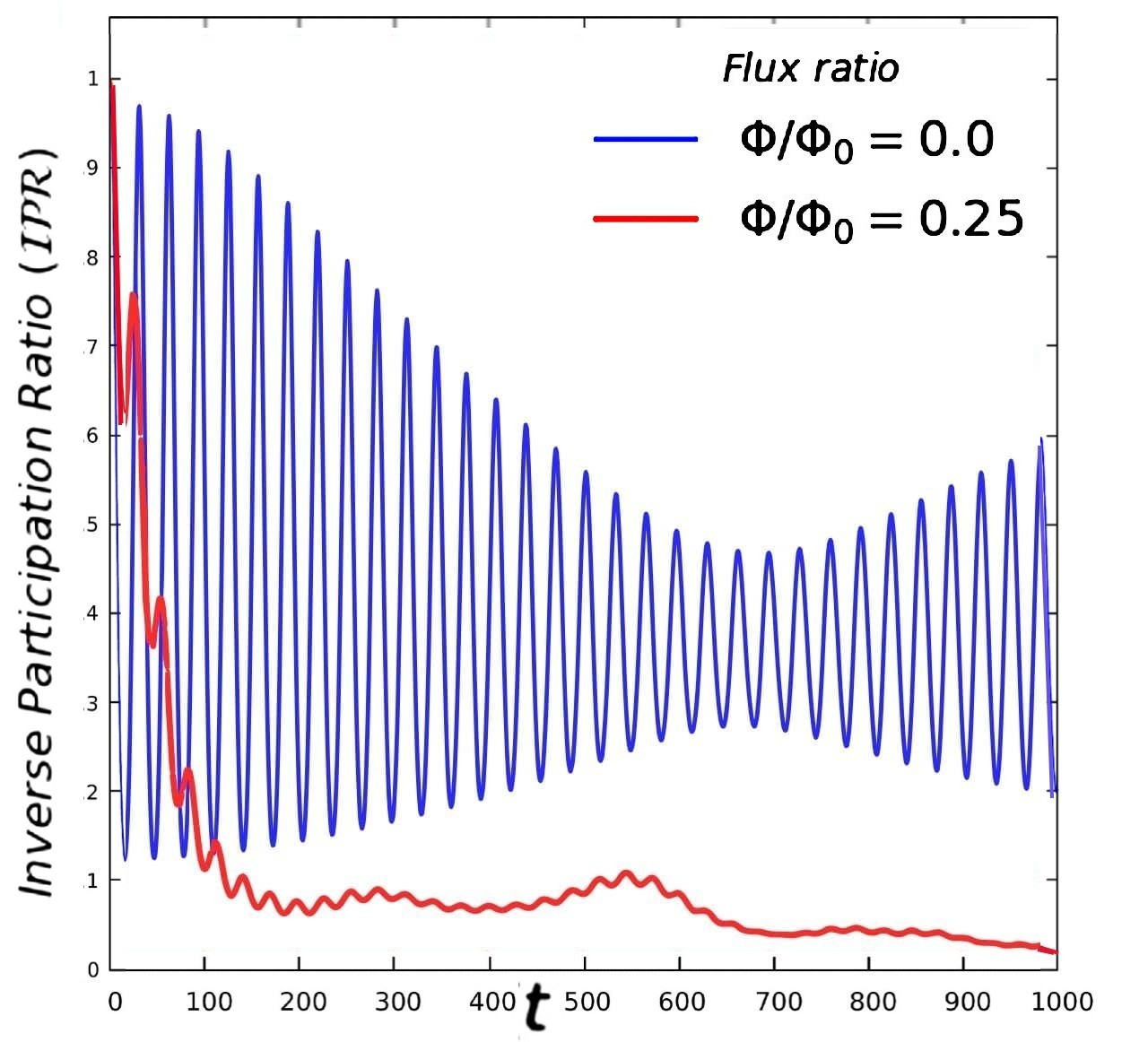}
\caption{IPR profile averaged over disordered systems of $1066$ sites for the zero flux and resonant flux configuration $\Phi=\Phi_0/4$. For both the curves we have set $V_1/V_2=\sqrt{2}$.}
\label{sdipr}
\end{figure}
%%%%%%%%%%%%%%%%%%%%%%%%%%%%%%%%%%%%%%%%%%%%%%%%

The next measure of investigation is the IPR, and we increase the size of the system to have $1000$ lattice points in total. We keep the hopping ratio $V_1/V_2=\sqrt{2}$ and release the wavefunction  from the same site for two different cases, viz, one with no flux in the plaquettes and one with a flux equal to $\Phi_0/4$.  The resulting IPRs for completely disordered arrangements of squares and dots are studied. The IPR's are averaged over ten disorder configurations and are shown in Fig.~\ref{sdipr}. In the absence of flux, the IPR shows an extreme oscillatory behaviour over the time period of study and never reaches the $1/N$ order of magnitude. This oscillatory behaviour can be accepted as a signature of significant backscattering of the wave function as it `proceeds' in the direction of the growth. Localization is apparent here. The rises are due to the spread of the wavefunction, thereby bringing more lattice sites into play, while the dips are due to the scattering which causes the wavefunction to track back and localize before spreading out again. The trend of any delocalization is completely obscure and the system undoubtedly favours localization. As the flux settles to the value $\Phi=\Phi_{0}/4$,
the IPR quickly drops from a value of `$1$' to a value very close to zero as time flows, and eventually gets saturated at a pretty low value. These features are all evidences of a {\it complete delocalization} in the system triggered by tuning the field to a special value. The averaging over different disorder arrangements makes our claims robust.
\section{\label{sec:level1}Summary and Conclusion\protect  }
In the present work we have substantiated an earlier observation ~\cite{biplab1} that, in certain class of geometrically disordered arrangement of scatterers, with a minimal quasi-one dimensionality introduced by hand, a numerical correlation between the hopping parameters can render all the single particle states completely extended, by an extensive study of the quantum dynamics. In particular, we have studied the mean square displacement, the temporal autocorrelation function and the time development of the average inverse participation ratio. The studies encompass a deterministic quasiperiodic sequence of triangles and dots as the two basic building elements of a chain, as well as a completely random array of a square geometry and dots lying on a one dimensional backbone. Even in a completely disordered lattices with spatial disorder, the special correlations are found to lead to a total breakdown of Anderson localization. These correlations generally cause commutation of all possible transfer matrices describing the system, leading to a visualization of an isomorphic periodic structure and ballistic transport,  or in certain cases, they cause diffusive or subdiffusive transport. The most important idea put forward in such a dynamical study is that, the observed localization - delocalization transition can be completely monitored by an external magnetic field. This makes the study of such physical phenomena important from the standpoint of designing a switching device. This is an interesting result,in the sense that it gives us a level of external control over the lattice. The studies bring in a challenge to examine an equivalent situation in exotic lattice structures that are nowadays built via interference of coherent lasers, or, ultra-fast laser writing techniques. By controlling the frequency and spatial orientation of the lasers, one can tailor similar lattices and experimentally verify such claims.

However, our results so far are applicable to single particle systems only.
For many body systems, the Hamiltonian changes to include interactions terms, viz the many body Hubbard model. In that model, for weak interactions, the ground state is basically a superfluid. The bosons or fermions in the many body model
altogether form a common macroscopic wavefunction
,and are totally delocalized
throughout the lattice.  If the interactions become  strong, the
ground state is a ‘Mott insulator’ where the
particles becomes localized to individual lattice
sites. In that state, the number of atoms in each
site is invariant, resulting in the disappearance of superfluidity.
It will be interesting to see whether just by maintaining the specific correlation relations in the lattice, a delocalised state can be generated or not.

\begin{acknowledgments}
AM is supported by the MSc studentship at IACS, sponsored by the Department of Science
and Technology, Govt. of India.
\end{acknowledgments}


\begin{thebibliography}{99}
\bibitem{anderson} P. W. Anderson, Phys. Rev. \textbf{109}, 1492 (1958).
\bibitem{abrahams} E. Abrahams, P. W. Anderson, D. C. Licciardello, and T. V. Ramakrishnan, Phys. Rev. Lett. \textbf{42}, 673 (1979).
\bibitem{borland} B. E. Borland, Proc. R. Soc. London \textbf{A 529}, 274 (1963).
\bibitem{wiersma} D. S. Wiersma, P. Bertolini, A. Lagendijk, and R. Righini, Nature \textbf{390}, 671 (1997).
\bibitem{cao} H. Cao, Y. G. Zhao, S. T. Ho, E. W. Seelig, Q. H. Wang, and R. P. H. Chang, Phys. Rev. Lett. \textbf{82}, 2278 (1999).
\bibitem{storzer} M. St\"{o}rzer, P. Gross, Christof M. Aegerter, and G. Maret
Phys. Rev. Lett. \textbf{96}, 063904 (2006). 
\bibitem{lahini} Y. Lahini, A. Avidan, F. Pozzi, M. Sorel, R. Morandotti, D. N. Christodoulides, and Y. Silberberg, Phys. Rev. Lett. \textbf{100}, 013906 (2008).
\bibitem{aspect} J. Billy, V. Josse, Z. Zuo, A. Bernard, B. Hambrecht, P. Lugan, D. Clément, L. Sanchez-Palencia, P. Bouyer, and A. Aspect, Nature \textbf{453}, 891 (2008).
\bibitem{roati} G. Roati, C. D’Errico, L. Fallani, M. Fattori, C. Fort, M. Zaccanti, G. Modugno, M. Modugno, and  M. Inguscio,  
Nature \textbf{453}, 895 (2008).
\bibitem{dunlap} D. H. Dunlap, H. -L. Wu, and P. W. Phillips, Phys. Rev. Lett. \textbf{65}, 88 (1990).
\bibitem{phillips} P. Phillips and H. -L. Wu, Science \textbf{252}, 1805 (1991).
\bibitem{francisco} V. Bellani, E. Diez, R. Hey, L. Toni, L. Tarricone, G. B. Parravicini, F. Dominguez-Adame, and R. G\'{o}mez-Alcal\'{a}, Phys. Rev. Lett. \textbf{82}, 2159 (1999).
\bibitem{naether} U. Naether, S. St\"{u}tzer, R. A. Vicencio, M. I. Molina, A. T\"{u}nnermann, S. Nolte, T. Kottos, D. N. Christodoulides, and A. Szameit, New. J. Phys. \textbf{15}, 013045 (2013).
\bibitem{macia} E. Maci\'{a} and F. Dominguez-Adame, Phys. Rev. Lett. \textbf{76}, 2957 (1996).
\bibitem{arunava1} A. Chakrabarti, S. N. Karmakar, and R. K. Moitra, Phys. Rev. Lett. \textbf{74}, 1403 (1995).
\bibitem{arunava2} A. Chakrabarti, S. N. Karmakar, and R. K. Moitra, Phys. Rev. B \textbf{50}, 13276 (1994).
\bibitem{arunava3} A. Chakrabarti and B. Bhattacharyya, Phys. Rev. B \textbf{56}, 13768 (1997).
\bibitem{biplab1} B. Pal, S. K. Maiti and A. Chakrabarti, Europhys. Lett. \textbf{102}, 17004 (2013).
\bibitem{biplab2} A. Nandy, B. Pal, and A. Chakrabarti, Europhys. Lett. \textbf{115}, 37004 (2016).
\bibitem{amrita} A. Mukherjee, A. Chakrabarti, and R. A. R\"{o}mer, Phys. Rev. B \textbf{98}, 075415 (2018).
\bibitem{katsanos} D. E. Katsanos, S. N. Evangelou, and S. J. Xiong, Phys. Rev. B \textbf{51}, 895 (1995).
\bibitem{arias} S. De Toro Arias and J. M. Luck, J. Phys. A: Math. Gen. \textbf{31}, 7699 (1998).
\bibitem{zhong} J. X. Zhong and R. Mosseri, J. Phys.: Condens. Matter \textbf{7}, 8383 (1995).
\bibitem{thiem1} S. Thiem, M. Schreiber, and U. Grimm, Phys. Rev. B \textbf{80}, 214203 (2009).
\bibitem{thiem2} S. Thiem and M. Schreiber, Phys. Rev. B \textbf{86}, 224205 (2012).
\bibitem{thiem3} S. Thiem, Phil. Mag. \textbf{95}, 1233 (2015).
\bibitem{sebabrata} S. Mukherjee, Marco Di Liberto, P. Öhberg, R. R. Thomson, and  N. Goldman, Phys. Rev. Lett. \textbf{121}, 075502 (2018).
\bibitem{roati} G. Roati, C. D'Errico, L. Fallani, M. Fattori, C. Fort, M. Zaccanti, G. Modugno, M. Modugno, and M. Inguscio, 
Nature \textbf{453}, 895 (2008).
\bibitem{kohmoto} M. Kohmoto, B. Sutherland, and C. Tang, Phys. Rev. B \textbf{35},  1020 (1987).
\end{thebibliography}
\end{document}